\documentclass[prc,superscriptaddress,unsortedaddress,twocolumn,showpacs]{revtex4}
\usepackage{graphicx}
\usepackage{amsmath}
\usepackage{amssymb}
\usepackage{times}
\usepackage{bm}
\usepackage{ulem}
\usepackage{color}

\def\dfrac#1#2{{\displaystyle\frac{#1}{#2}}}
\newcommand{\beq}{\begin{equation}}
\newcommand{\eeq}{\end{equation}}
\newcommand{\bea}{\begin{eqnarray}}
\newcommand{\eea}{\end{eqnarray}}

\def\non{\nonumber}
\def\la{\mathrel{\mathpalette\fun <}}
\def\ga{\mathrel{\mathpalette\fun >}}
\def\fun#1#2{\lower3.6pt\vbox{\baselineskip0pt\lineskip.9pt
\ialign{$\mathsurround=0pt#1\hfil##\hfil$\crcr#2\crcr\sim\crcr}}}

\begin{document}

\title{
Asymmetry of the parallel momentum distribution of ($p,pN$) reaction residues
}

\author{Kazuyuki Ogata}
\email[]{kazuyuki@rcnp.osaka-u.ac.jp}
\affiliation{Research Center for Nuclear Physics (RCNP), Osaka
University, Ibaraki 567-0047, Japan}

\author{Kazuki Yoshida}
\affiliation{Research Center for Nuclear Physics (RCNP), Osaka
University, Ibaraki 567-0047, Japan}

\author{Kosho Minomo}
\affiliation{Research Center for Nuclear Physics (RCNP), Osaka
University, Ibaraki 567-0047, Japan}

\date{\today}

\begin{abstract}
The parallel momentum distribution (PMD) of the residual nuclei of the
$^{14}$O($p,pn$)$^{13}$O and $^{14}$O($p,2p$)$^{13}$N reactions
at 100 and 200~MeV/nucleon in inverse kinematics is investigated
with the framework of the distorted wave impulse approximation.
The PMD shows an asymmetric shape characterized by a steep fall-off
on the high momentum side and a long-ranged tail on the low momentum
side. The former is found to be due to the phase volume effect reflecting
the energy and momentum conservation, and the latter is to the
momentum shift of the outgoing two nucleons inside an
attractive potential caused by the residual nucleus.
Dependence of these effects on the nucleon separation energy
of the projectile and the incident energy is also discussed.
\end{abstract}

% 24.10.Eq    Coupled-channel and distorted-wave models
% 25.60.Gc    Breakup and momentum distributions

\maketitle

\section{Introduction}
\label{sec1}

The single-particle (s.p.) nature is one of the most important
properties of nuclei. Since 90's, triggered by the invention of
radioactive isotope (RI) beam technology, intensive studies on the s.p. structure
of unstable nuclei have been done; see, for a review, Ref.~\cite{Tan13}.
Many experiments of one- or two-nucleon removal processes were
performed~\cite{Kel95,Nav98,Neg99,Sau00,Mad01,End02,Baz03,Gad04,Sau04,%
Gad05,Gad08,Baz09,Gri12,Str14,TG14}
and the parallel momentum distribution (PMD) of
the residual nucleus B has widely been used to determine the
s.p. structure of the incident particle A, i.e., the orbital angular
momentum, the s.p. energy, and the spectroscopic factor of
the nucleon(s) inside the nucleus A.

The removal reaction does not specify the final state of the
target nucleus T, and it is not trivial to apply a direct reaction theory
to such an inclusive process. The Glauber model~\cite{Gla59,HM85,Hen96,TB06}
is one of the most successful models to study the nucleon removal process;
the eikonal and adiabatic approximations allow one to treat
the scattering of each constituent of A off T separately, and
the generalized unitarity of the scattering matrix elements of the removed
nucleon(s) gives a simple form of reaction observables for the
nucleon removal process.
Recently, the eikonal reaction theory (ERT)~\cite{Yah11} was developed
as an extension of the continuum-discretized coupled-channels method
(CDCC)~\cite{Kam86,Aus87,Yah12}.
Although ERT has successfully been applied to one- and two-neutron removal
processes~\cite{Min14}, formulation of the PMD and differential cross sections
with the ERT has not been completed.

Despite the great success of the Glauber model in extracting s.p.
information on unstable nuclei from reaction data,
the shape of the PMD calculated with the Glauber model
is restricted to being symmetric. On the other hand,
in some cases the observed PMD shows
quite large asymmetry~\cite{End02,Gad04,Gad05}.
Although this does not necessarily cause ambiguity of
the s.p. information extracted by the Glauber model, as discussed
in Ref.~\cite{Gad05}, it will be interesting and important
to clarify the mechanism of the asymmetry of the PMD.
It should be remarked that recently the transfer to the continuum (TC)
method~\cite{BB88,BB91}
was applied to the one-neutron removal process from $^{14}$O by
$^{9}$Be at 53~MeV/nucleon and reproduced quite well the
asymmetric shape of the PMD~\cite{Fla12}. At this moment, the TC method
is restricted to neutron removal processes.

As mentioned, it is difficult to describe an inclusive process, to which
huge numbers of the final states of T contribute.
On the other hand, the description of elastic breakup (EB), in which
T stays in the ground state in the final channel, is
well established. CDCC~\cite{Kam86,Aus87,Yah12},
the dynamical eikonal approximation (DEA)~\cite{Bay05,Gol06},
the Faddeev--Alt-Grassberger-Sandhas (Faddeev-AGS) theory~\cite{Fad60,Alt67} etc.
can be adopted; CDCC is even applicable to the nuclear and Coulomb breakup of
a three-body projectile by a target nucleus~\cite{Mat04,Mat10,Rod08,Rod09}.

The asymmetry of the PMD, its EB component in particular,
of $^{14}$C after the one-neutron removal from $^{15}$C by $^{9}$Be
at 54~MeV/nucleon was discussed in Ref.~\cite{Tos02} by means of CDCC.
It was concluded that the accurate treatment of three-body reaction
dynamics was essential to reproduce the asymmetry of the PMD of $^{14}$C.
Because of this finding, sometimes the asymmetry of a PMD is
regarded as a result of higher-order effects.
Discussion on the very large asymmetry found in
$^{9}$Be($^{46}$Ar,$^{45}$Ar\,$x$), where $x$
indicates that all other particles are not detected,
at 70~MeV/nucleon~\cite{Gad05}, however, has not been done.

In the present study, we focus on one-nucleon knockout processes
by a hydrogen target, i.e., ($p,pN$) reactions in inverse kinematics,
in which only the EB process occurs.
($p,pN$) reactions have been used to determine the s.p. structure
of nuclei; for reviews, see Refs.~\cite{JM66,JM73,Kit85}.
An important feature of
($p,pN$) reactions is that the energy and momentum transfer
($\omega$-$q$) as well as the angular momentum transfer $\Delta l$
is large in general. This makes the reaction mechanism rather simple,
and the distorted-wave impulse approximation (DWIA) has been
successful in describing ($p,pN$) reaction observables. Recently,
a comparison between the DWIA and the Faddeev-AGS theory was done
for $^{11}$Be($p,pn$)$^{10}$Be and  $^{12}$C($p,pn$)$^{11}$B,
and the results of the two methods were shown to agree very well
with each other
at proton energies above 100~MeV~\cite{Cre08,Cre14}. It should be noted that
the Glauber model~\cite{Gla59,HM85,Hen96,TB06},
which relies on the adiabatic approximation,
assumes small $\omega$-$q$ and is thus not suitable for describing
($p,pN$) reactions. CDCC does not use the adiabatic approximation
and is applicable to ($p,pN$) reactions~\cite{Kon09,Kon10,Oza11}.
However, the model space of CDCC required to describe ($p,pN$) reactions
is large mainly because of the large value of $\Delta l$.
Furthermore, for a knockout
process of a tightly bound nucleon, $\omega$-$q$ and $\Delta l$
are even larger, which will make CDCC unpractical.

In this paper we investigate the asymmetry of the PMD of
$^{14}$O($p,pN$) reactions in inverse kinematics at 100 and
200~MeV/nucleon. $^{14}$O has the neutron and proton separation energies
of 23.2~MeV and 4.63~MeV, respectively, and their large difference
is expected to give a quite different shape to the PMD of the
reaction residues, i.e., $^{13}$O and $^{13}$N.
The main purpose of the present study is to understand the mechanism
that gives the asymmetric shape of the PMD.
We adopt the DWIA with the eikonal approximation, the accuracy of which
is judged by comparison with the experimental data of the
triple differential cross section (TDX) of $^{12}$C($p,pN$)$^{11}$B
at 392~MeV~\cite{Nor05}.
Roles of the phase volume, which guarantees the energy and
momentum conservation, and the distortion of the outgoing two nucleons
by the residual nucleus are investigated separately.
Quite recently, in Ref.~\cite{Aum13} an eikonal DWIA model
was developed for describing the momentum distribution (MD)
of ($p,pN$) reaction resides.
The authors focused on reactions at around 500~MeV/nucleon, and
the asymmetry of the PMD, which is expected to appear at lower energies,
has not been discussed.

The construction of this paper is as follows. In Sec.~\ref{sec2}
formulation of the TDX and PMD with the eikonal DWIA is given.
In Sec.~\ref{sec3} first we show the accuracy of the
eikonal DWIA by comparing the result of the TDX of
$^{12}$C($p,pN$)$^{11}$B at 392~MeV with experimental data.
We then show the PMD of the $^{14}$O($p,pN$) reaction residues
at 100~MeV/nucleon. The phase volume effect on the high momentum
side and the distortion effect on the low momentum side
are discussed in detail. Results at 200~MeV/nucleon are also
investigated. Finally, we give a summary in Sec.~\ref{sec4}.

\section{Formalism}
\label{sec2}

We consider A$(p,2p)$B and A$(p,pn)$B reactions in inverse or normal
kinematics. We adopt the framework of the DWIA to calculate the TDX and
the PMD. Both observables are evaluated
in the frame in which A is at rest, i.e., the A-rest frame.
We refer to the proton in the initial channel as particle 0,
and the outgoing two nucleons to as particle 1 and 2.
The momentum (in the unit of $\hbar$) and the total energy of
particle $i$ $(=0$, 1, 2, or B) are denoted by
$\boldsymbol{K}_{i}$ and $E_i$, respectively;
$T_i$ represents the kinetic part of $E_i$ and
$\Omega_i$ is the solid angle of $\boldsymbol{K}_{i}$.
These quantities with and without the superscript A mean that they
are evaluated in the A-rest frame and the $p$-A center-of-mass (c.m.)
frame, respectively.
In the following equations, we adopt the relativistic kinematics for each
particle, i.e.,
$E_i^{\mathrm{A}}=\sqrt{(m_i c^2)^2+(\hbar c K_i^{\mathrm{A}})^2}$
and
$E_i=\sqrt{(m_i c^2)^2+(\hbar c K_i)^2}$,
with $m_i$ the rest mass of particle $i$, are used.
The Lorentz transformation is adopted to relate the
four-dimensional momenta $(\hbar {\bm K}_i^{\mathrm{A}},E_i^{\mathrm{A}}/c)$
and $(\hbar {\bm K}_i,E_i/c)$.

The antisymmetrization between particles 1 and 2 is understood to
be taken into account in the nucleon-nucleon ($NN$) transition matrix
$t_{NN}$, which is treated approximately by using
\beq
\dfrac{(m_N /2)^2}{(2\pi \hbar^2)^2}
\left|t_{NN}\right|^2
\approx
\bar{\sigma}_{NN}.
\label{signn}
\eeq
The nucleon mass is denoted by $m_N$ and $\bar{\sigma}_{NN}$ is
the $NN$ elastic differential cross section averaged over incident
energies and scattering angles relevant to the knockout process
considered.
Equation~(\ref{signn}) is in fact based on the following three
approximations. First, $t_{NN}$ is evaluated with the $NN$ asymptotic
kinematics~\cite{CR77}. Second, the off-the-energy-shell $t_{NN}$
in the nuclear medium is approximated by an averaged on-shell
matrix element in free space. Then one can evaluate $|t_{NN}|^2$ by
the $NN$ differential cross section multiplied by a kinetic
constant. Third, we take an average of the $NN$ differential cross
section to have a one number $\bar{\sigma}_{NN}$ to be used in the
calculation of the knockout process.
Note that $\bar{\sigma}_{NN}$ depends on the reaction type, i.e.,
($p,2p$) or ($p,pn$), and the incident energy of the
knockout process.
These rather drastic approximations to $t_{NN}$ are examined by
comparing the calculated TDX with experimental data in Sec.~\ref{sec32}.

We make the eikonal approximation to the
distorted waves of particles 0, 1, and 2, as in Ref.~\cite{Aum13};
we further approximate the eikonal
wave function with the forward scattering assumption; see
Eqs.~(\ref{fnor}) and (\ref{finv}) below.

The TDX is then given by
\begin{equation}
\frac{d^{3}\sigma}{dE_{1}^{\mathrm{A}}d\Omega_{1}^{\mathrm{A}}d\Omega
_{2}^{\mathrm{A}}}
=
F_{\mathrm{kin}}C_0
\sum_{m}
\bar{\sigma}_{NN} \left(  2\pi\right)  ^{2}
\left|
\bar{T}_{\boldsymbol{K}_{N},K_{0}K_{1}K_{2}}^{nljm}
\right|^2,
\label{tdx}
\end{equation}%
where
\begin{equation}
F_{\mathrm{kin}}\equiv
J_{\mathrm{A}}\frac{K_{1}K_{2}E_{1}E_{2}}{\hbar^{4}c^{4}}\left[
1+\frac{E_{2}}{E_{\mathrm{B}}}+\frac{E_{2}}{E_{\mathrm{B}}}\frac{\boldsymbol{K}_{1}%
\cdot\boldsymbol{K}_{2}}{K_{2}^{2}}\right]  ^{-1}
\end{equation}%
with $J_{\mathrm{A}}$ the Jacobian for the transformation from
the $p$-A c.m. frame to the A-rest frame, and
\begin{equation}
C_{0}=\frac{E_0^{\mathrm{A}}}{(\hbar c)^2 K_0^{\mathrm{A}}}
\frac{1}{\left(  2l+1\right)  }\frac
{4\hbar^{4}}{\left(  2\pi\right)  ^{3}m_{N}^{2}}.
\end{equation}
The reduced transition amplitude is given by
\bea
\bar{T}_{\boldsymbol{K}_{N},K_{0}K_{1}K_{2}}^{nljm}
&=&
\int
db\,bJ_{m}\left( K_{Nb}b \right)  \,dz\boldsymbol{\,}%
e^{-iK_{Nz}z}
\non \\
&&\times
F_{K_{0}K_{1}K_{2}}\left(  b,z\right)
\varphi_{nlj}\left(  R\right)  \bar{P}_{lm}\left(  \cos\theta_{R}\right),
\non \\
\label{tmat}
\eea
where $n$, $l$, and $j$ are, respectively, the principal quantum
number, the orbital angular momentum, and the total spin of the
s.p. orbit of the nucleon in A; $m$ is the third
component of $j$. $\varphi_{nlj}$ is the radial part of the s.p.
wave function, and $\bar{P}_{lm}$ is defined through the spherical harmonics
$Y_{lm}$ by
\beq
\bar{P}_{lm}(\cos\theta_{R})=Y_{lm}(\hat{\boldsymbol{R}}) e^{-im \phi_R},
\eeq
where $\hat{\boldsymbol{R}}$, $\theta_{R}$, and $\phi_{R}$ are
the solid, polar, and azimuthal angles of ${\bm R}$,
respectively. The $z$-axis is taken to be the direction of the
incident particle, i.e., $p$ (A) in normal (inverse) kinematics,
and $b$ is the length of the projected vector $\boldsymbol{b}$ of
$\boldsymbol{R}$ on the plane perpendicular to the $z$-axis.
$J_m$ is the Bessel function of the first kind.

The missing momentum $\boldsymbol{K}_{N}$ that plays a central role
in $(p,pN)$ reactions is defined by
\bea
\boldsymbol{K}_{N}&=&
\boldsymbol{K}_{1}+\boldsymbol{K}_{2}
-\dfrac{A-1}{A}\boldsymbol{K}_{0}=
-\boldsymbol{K}_{\mathrm{B}}
-\dfrac{A-1}{A}\boldsymbol{K}_{0} \non \\
&\equiv&
K_{Nz}{\bm e}_z + K_{Nb}{\bm e}_{\bm b},
\eea
where
$A$ is the mass number of A and
${\bm e}_z$ (${\bm e}_{\bm b}$) is  the unit vector for the
direction of $z$ (${\bm b}$).
It should be noted that with high accuracy one can find
$\boldsymbol{K}_{N}\approx - \boldsymbol{K}_{\mathrm{B}}^{\mathrm{A}}$.
If B is assumed to be a spectator in the ($p,pN$) process,
$\boldsymbol{K}_{N}$ can therefore be interpreted as the momentum of the struck
nucleon in the A-rest frame before the $NN$ collision.

In the present eikonal DWIA, the distortion effects for particle 0, 1,
and 2 are aggregated into the so-called distorted-wave factor
$F_{K_{0}K_{1}K_{2}}$ given by
\bea
F_{K_{0}K_{1}K_{2}}\left(  b,z\right)
&=&
\exp\left[  \frac
{1}{i\hbar v_{0}}\int_{-\infty}^{z}U_{0}\left(  b,z^{\prime}\right)
dz^{\prime}\right]
\non \\
&&\times
\exp\left[  \frac{1}{i\hbar v_{1}}\int_{z}^{\infty}%
U_{1}\left(  b,z^{\prime}\right)  dz^{\prime}\right]
\non \\
&&\times
\exp\left[  \frac
{1}{i\hbar v_{2}}\int_{z}^{\infty}U_{2}\left(  b,z^{\prime}\right)
dz^{\prime}\right]
\label{fnor}
\eea
in normal kinematics and by
\bea
F_{K_{0}K_{1}K_{2}}\left(  b,z\right)
&=&
\exp\left[  \frac
{1}{i\hbar v_{0}}\int_{z}^{\infty}U_{0}\left(  b,z^{\prime}\right)
dz^{\prime}\right]
\non \\
&&\times
\exp\left[  \frac{1}{i\hbar v_{1}}\int_{-\infty}^{z}%
U_{1}\left(  b,z^{\prime}\right)  dz^{\prime}\right]
\non \\
&&\times
\exp\left[  \frac
{1}{i\hbar v_{2}}\int_{-\infty}^{z}U_{2}\left(  b,z^{\prime}\right)
dz^{\prime}\right]
\label{finv}
\eea
in inverse kinematics. $U_i$ ($i=0$, 1, 2) is
the distorting potential for particle $i$ and
$v_i$ is its velocity.

The MD of B is defined by
\bea
\frac{d\sigma}{d\boldsymbol{K}_{\mathrm{B}}^{\mathrm{A}}}
&=&
C_{0}
\int
d{\bm K}_{1}^{\mathrm{A}}d{\bm K}_{2}^{\mathrm{A}}
\eta_{\mathrm{M\phi l}%
}^{\mathrm{A}}
\delta(E_f^{\mathrm{A}}-E_i^{\mathrm{A}})
\delta({\bm K}_f^{\mathrm{A}}-{\bm K}_i^{\mathrm{A}})
\non \\
&&\times
\bar{\sigma}_{NN}\sum_{m}\left(  2\pi\right)^2
\left|
\bar{T}_{\boldsymbol{K}_{N},K_{0}K_{1}K_{2}}^{nljm}
\right|^2,
\label{pmddef}
\eea
where
\beq
\eta_{\mathrm{M\phi l}%
}^{\mathrm{A}}=\frac{E_{1}E_{2}E_{\mathrm{B}}}{E_{1}^{\mathrm{A}}E_{2}^{\mathrm{A}%
}E_{\mathrm{B}}^{\mathrm{A}}}.
\eeq
The total energy (momentum) of the reaction system in the A-rest frame
in the initial and final channels is denoted by
$E_i^{\mathrm{A}}$ (${\bm K}_i^{\mathrm{A}}$)
and $E_f^{\mathrm{A}}$ (${\bm K}_f^{\mathrm{A}}$), respectively.
Equation~(\ref{pmddef}) can be reduced to
\bea
\frac{d\sigma}{d\boldsymbol{K}_{\mathrm{B}}^{\mathrm{A}}}
&=&
C_{0}
\frac{1}{\left(  \hbar c\right)  ^{2}Q^{\mathrm{A}}}\int
dK_{1}^{\mathrm{A}}\,K_{1}^{\mathrm{A}}E_{2}^{\mathrm{A}}
d\varphi_{1Q}^{\mathrm{A}}\,\eta_{\mathrm{M\phi l}}^{\mathrm{A}}
\non \\
&&\times
\bar{\sigma}_{NN}\sum_{m}\left(  2\pi\right)^2
\left|
\bar{T}_{\boldsymbol{K}_{N},K_{0}K_{1}K_{2}}^{nljm}
\right|^2,
\label{pmd2}
\eea
where ${\bm K}_2^{\mathrm{A}}$ is understood to be fixed at
${\bm K}_0^{\mathrm{A}}-{\bm K}_1^{\mathrm{A}}-
{\bm K}_{\mathrm{B}}^{\mathrm{A}}$ by the momentum conservation.
We measure the solid angle of
particle 1 with respect to
${\bm Q}^{\mathrm{A}}\equiv {\bm K}_0^{\mathrm{A}}
- {\bm K}_{\mathrm{B}}^{\mathrm{A}}$ as
\beq
d{\bm K}_1^{\mathrm{A}}=(K_1^{\mathrm{A}})^2 d K_1
d (\cos{\theta_{1Q}^{\mathrm{A}}})
d \varphi_{1Q}^{\mathrm{A}}.
\eeq
In Eq.~(\ref{pmd2}) $\cos{\theta_{1Q}^{\mathrm{A}}}$ has been fixed
at a value that satisfies the energy conservation; this value
as well as the lower and upper limits of $K_1^{\mathrm{A}}$
is given analytically. The PMD is given by integrating the MD
over the absolute value of the $b$ component
of ${\bm K}_{\mathrm{B}}^{\mathrm{A}}$:
\beq
\frac{d\sigma}{d {K}_{\mathrm{B}z}^{\mathrm{A}}}
=
\int d {K}_{\mathrm{B}b}^{\mathrm{A}}\, K_{\mathrm{B}b}^{\mathrm{A}}
\frac{d\sigma}{d\boldsymbol{K}_{\mathrm{B}}^{\mathrm{A}}}.
\label{pmdfin}
\eeq
When the theory is compared with PMD data integrated over the azimuthal angle
of $\boldsymbol{K}_{\mathrm{B}}^{\mathrm{A}}$, Eq.~(\ref{pmdfin})
must be multiplied by $2\pi$. For reactions in inverse kinematics,
the A-rest frame is different from the laboratory frame (L-frame).
Then
\beq
\frac{d\sigma}{d\boldsymbol{K}_{\mathrm{B}}^{\mathrm{L}}}
=
\dfrac{E_{\mathrm{B}}^{\mathrm{A}}}{E_{\mathrm{B}}^{\mathrm{L}}}
\frac{d\sigma}{d\boldsymbol{K}_{\mathrm{B}}^{\mathrm{A}}}
\eeq
can be used for comparison with experimental data, if necessary. The
superscript L indicates the L-frame.

If plane wave impulse approximation (PWIA) is adopted, further
simplification of Eq.~(\ref{pmd2}) can be done:
\beq
\frac{d\sigma^{\mathrm{PW}}}{d\boldsymbol{K}_{\mathrm{B}}^{\mathrm{A}}}
=
\bar{\rho}_{{\bm K}_{\mathrm{B}}^{\mathrm{A}}}
\mathfrak{F}_{{\bm K}_{\mathrm{B}}^{\mathrm{A}}},
\label{pmdpw}
\eeq
where the phase volume $\bar{\rho}_{{\bm K}_{\mathrm{B}}^{\mathrm{A}}}$ is given by
\beq
\bar{\rho}_{{\bm K}_{\mathrm{B}}^{\mathrm{A}}}=
\frac{1}{\left(  \hbar c\right)  ^{2}Q^{\mathrm{A}}}\int
dK_{1}^{\mathrm{A}}\,K_{1}^{\mathrm{A}}E_{2}^{\mathrm{A}}
d\varphi_{1Q}^{\mathrm{A}}\,\eta_{\mathrm{M\phi l}}^{\mathrm{A}}
\label{pvr}
\eeq
and the effective s.p. MD is defined by
\beq
\mathfrak{F}_{{\bm K}_{\mathrm{B}}^{\mathrm{A}}}
=
C_{0}
\bar{\sigma}_{NN}
\sum_{m}
\left|
\tilde{\phi}_{nljm}({\bm K}_N)
\right|^2
\eeq
with
\beq
\tilde{\phi}_{nljm}({\bm K}_N)
=
\int d {\bm R}\,
e^{-i{\bm K}_N \cdot {\bm R}}
\varphi_{nlj}(R) Y_{lm}(\hat{\bm R}).
\label{phitil}
\eeq
Thus $\bar{\rho}_{{\bm K}_{\mathrm{B}}^{\mathrm{A}}}$
and $\mathfrak{F}_{{\bm K}_{\mathrm{B}}^{\mathrm{A}}}$ are factorized.
{Equations (\ref{pmdpw})--(\ref{phitil}) are used
in Sec.~\ref{sec33} to see the phase volume effect on the PMD.}

In the actual calculation Eq.~(\ref{pvr}) is used, i.e., the energy and
momentum conservation based on the relativistic kinematics is
taken into account. For an interpretation of the numerical result, however,
the following nonrelativistic expression of
$\bar{\rho}_{{\bm K}_{\mathrm{B}}^{\mathrm{A}}}$ will be helpful:
\beq
\bar{\rho}_{{\bm K}_{\mathrm{B}}^{\mathrm{A}}}^{\mathrm{NR}}
\equiv=
\frac{\pi m_N}{\hbar^2}
\sqrt{
({\bm K}_0^{\mathrm{A}} + {\bm K}_{\mathrm{B}}^{\mathrm{A}})^2
-\dfrac{2A}{A-1}(K_{\mathrm{B}}^{\mathrm{A}})^2
-\bar{S}_N
},
\label{pv}
\eeq
where
$\bar{S}_{N}\equiv4m_N S_N/\hbar^2$
with $S_N$ the nucleon separation energy of A.
Although it is rather trivial, the argument of the square root
in Eq.~(\ref{pv}) is understood to be not negative; this condition
determines, within the nonrelativistic kinematics, the allowed
region of ${\bm K}_{\mathrm{B}}^{\mathrm{A}}$
for satisfying the energy and momentum conservation.
A similar discussion can be done in the relativistic kinematics.
However, the functional form of
$\bar{\rho}_{{\bm K}_{\mathrm{B}}^{\mathrm{A}}}$
is much more complicated than Eq.~(\ref{pv}).

\section{Results and discussion}
\label{sec3}

\subsection{Numerical inputs}
\label{sec31}

We use the s.p. potential of Bohr and Mottelson~\cite{BM69}
for the nucleon inside the nucleus. The depth of its central part
is changed so as to reproduce $S_N$.
For the nucleon distorting potential below (above) 200~MeV,
we adopt the parameter set of Koning and Delaroche~\cite{KD03}
(Dirac phenomenology~\cite{Ham90}); its energy dependence is
explicitly taken into account.
We multiply each of the distorted waves by
the Perey factor~\cite{PB62}
$F_{\mathrm{Per}}(R)=[1-\mu \beta^2/(2\hbar^2)U(R)]^{-1/2}$,
where $\mu$ is the reduced mass between the scattering two particles,
to include the effect of the nonlocality of the distorting potential;
the range $\beta$ of nonlocality is chosen to be 0.85~fm.
The parametrization of Franey and Love~\cite{FL85}
for $t_{NN}$ is used.

\subsection{TDX and PMD calculated with eikonal DWIA}
\label{sec32}

%%%%%%%%%%%%%%%%%%%%%%%
%%%  Figure 1
%%%%%%%%%%%%%%%%%%%%%%%
\begin{figure}[htbp]
\begin{center}
 \includegraphics[width=0.45\textwidth,clip]{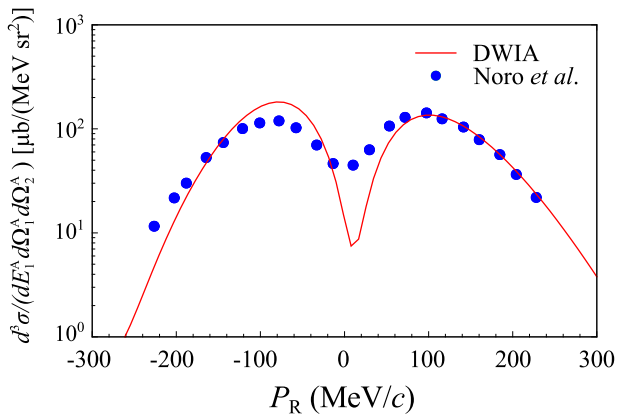}
 \caption{(Color online) Triple differential cross section for
 the $^{12}$C($p,2p$)$^{11}$B reaction at 392~MeV in normal kinematics.
 The definition of the horizontal axis and detailed kinematical conditions
 for the outgoing protons are given in  the text.
 The experimental data are taken from Ref.~\cite{Nor05}.
}
 \label{fig1}
\end{center}
\end{figure}
First we test the accuracy of the eikonal DWIA model described
in Sec.~\ref{sec2}. We calculate the TDX for the $^{12}$C($p,2p$)$^{11}$B
reaction at 392~MeV in normal kinematics with Eq.~(\ref{tdx}).
We follow the Madison convention
and the kinematics of the outgoing two protons is fixed as
$T_1=250$ (MeV), $\theta_1=32.5^\circ$, $\phi_1=0^\circ$, and
$\phi_2=180^\circ$; this means that the energy transfer
$\hbar \omega$ is 142~MeV and the momentum transfer $q$ is 2.59~fm$^{-1}$.
We assume that the $0p3/2$ proton in $^{12}$C is knocked out.
In Fig.~\ref{fig1} we show the result of the TDX and
the experimental data~\cite{Nor05}. We use the spectroscopic factor
${\cal S}=1.72$ determined by the ($e,e'p$) experiment~\cite{Ste88}.
The horizontal axis is defined by
\beq
P_{\mathrm{R}} = \hbar K_{\mathrm{B}}^{\mathrm{A}}
\dfrac{K_{\mathrm{B}z}^{\mathrm{A}}}{|K_{\mathrm{B}z}^{\mathrm{A}}|}.
\eeq
One sees the calculation reproduces the data very well,
which suggests the success of the present reaction model,
i.e., the eikonal DWIA with the forward scattering assumption
and the use of the averaged
$NN$ cross section in free space.
It should be noted that the undershooting at around
$P_{\mathrm{R}}=0$, which corresponds to the so-called
quasi-free condition (QFC),
is mainly because the experimental data have been
integrated over
$d\Omega_{1}^{\mathrm{A}}$ and $d\Omega_{2}^{\mathrm{A}}$
in the range of the resolution of the detectors~\cite{Nor05}.
The calculated TDX reproduces the data up to $|P_R|\sim 200$~MeV/$c$,
i.e., well away from the QFC point. This will be due in part to
the kinematical condition corresponding to large $\omega$-$q$.

%%%%%%%%%%%%%%%%%%%%%%%
%%%  Figure 2
%%%%%%%%%%%%%%%%%%%%%%%
\begin{figure}[htbp]
\begin{center}
 \includegraphics[width=0.45\textwidth,clip]{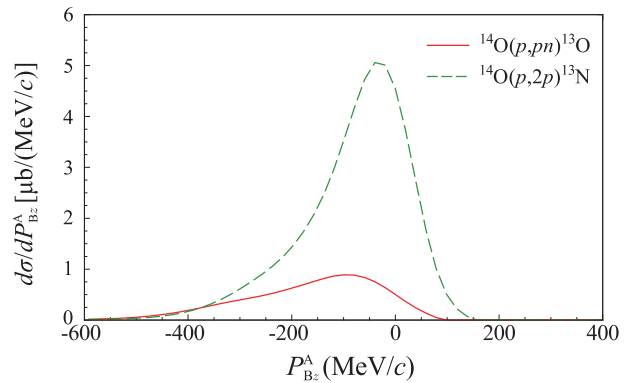}
 \caption{(Color online) Parallel momentum distribution of
 the $^{14}$O($p,pn$)$^{13}$O (solid line)
 and $^{14}$O($p,2p$)$^{13}$N (dashed line)
 reaction residues at 100~MeV/nucleon in inverse kinematics.
}
 \label{fig2}
\end{center}
\end{figure}
Next we show in Fig.~\ref{fig2} the PMD of the residual nuclei of
the $^{14}$O($p,pn$) (solid line) and $^{14}$O($p,2p$) (dashed line)
reactions at 100~MeV/nucleon, as a function of
$P_{\mathrm{B}z}^{\mathrm{A}} = \hbar K_{\mathrm{B}z}^{\mathrm{A}}$.
For the former (latter) the $0p3/2$ neutron ($0p1/2$ proton)
with $S_N=23.2$~MeV (4.63~MeV) is
assumed to be knocked out by the target proton;
from now on we always use ${\cal S}=1$.
In both reactions the PMD shows clear
asymmetry. For more quantitative discussion, we divide the FWHM $\Gamma$
into two parts corresponding to
the low ($\Gamma_{\mathrm{L}}$) and high
($\Gamma_{\mathrm{H}}$) momentum sides with respect to the peak
position $P_{\mathrm{cen}}$:
\beq
\Gamma = \Gamma_{\mathrm{L}} + \Gamma_{\mathrm{H}}.
\eeq
The asymmetry $A_{\Gamma}$
is defined by $\Gamma_{\mathrm{L}}/\Gamma_{\mathrm{H}}$.
%
%%%%%%%%%%%%%%%%%%%%%%%
%%%  Table 1
%%%%%%%%%%%%%%%%%%%%%%%
\begin{table}[hptb]
\caption{$P_{\mathrm{cen}}$, $\Gamma$, $\Gamma_{\mathrm{L}}$,
$\Gamma_{\mathrm{H}}$, and $A_{\Gamma}$ for the $^{14}$O($p,pN$)
reaction residues at 100~MeV/nucleon.}
\label{tab1}
\begin{tabular}{cccccc}
\hline
\hline
nucleus  & $P_{\mathrm{cen}}$~(MeV/$c$) & $\Gamma$~(MeV/$c$)
         & $\Gamma_{\mathrm{L}}$~(MeV/$c$)
         & $\Gamma_{\mathrm{H}}$~(MeV/$c$) & $A_{\Gamma}$ \\
\hline
$^{13}$O  & $-92$  & 266 & 182 &  84 & 2.17 \\
$^{13}$N  & $-31$  & 178 & 104 &  74 & 1.41 \\
%% $^{13}$O  & $-53$  & 278 & 168 & 110 & 1.53 \\
%% $^{13}$N  & $-20$  & 191 & 113 &  78 & 1.45 \\
%% $^{31}$Ne & $-0.7$ &  52 &  32 &  30 & 1.08 \\
\hline
\hline
\end{tabular}
\end{table}
The values of $P_{\mathrm{cen}}$, $\Gamma$, $\Gamma_{\mathrm{L}}$,
$\Gamma_{\mathrm{H}}$, and $A_{\Gamma}$ are listed in Table~\ref{tab1}.
One sees that the shift of the peak position
from the origin, $|P_{\mathrm{cen}}|$, of $^{13}$O is three times
as large as that of $^{13}$N.
Another finding is that the two residues have similar values of
$\Gamma_{\mathrm{H}}$, whereas $\Gamma_{\mathrm{L}}$ of them
are quite different from each other.
In Secs.~\ref{sec33} and \ref{sec34} we investigate these features
of the PMD in more detail.

\subsection{Phase volume effect on the high momentum side}
\label{sec33}

%%%%%%%%%%%%%%%%%%%%%%%
%%%  Figure 3
%%%%%%%%%%%%%%%%%%%%%%%
\begin{figure}[htbp]
\begin{center}
 \includegraphics[width=0.45\textwidth,clip]{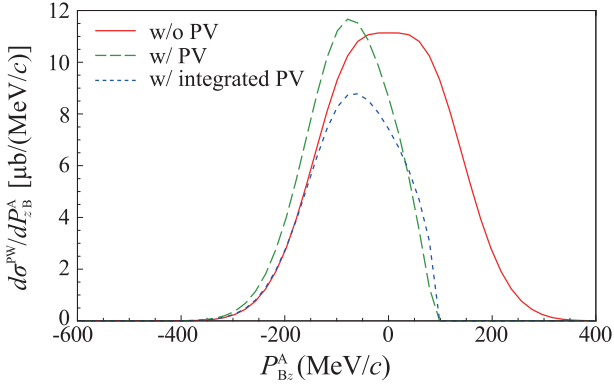}
 \caption{(Color online) Parallel momentum distribution of
 the $^{14}$O($p,pn$)$^{13}$O
 reaction residue at 100~MeV/nucleon in inverse kinematics.
 The solid (dashed) line shows the result of the PWIA without (with)
 taking into account the phase volume. The dotted line is
 the same as the dashed line but an averaged phase volume
 is used (see the text for detail).
}
 \label{fig3}
\end{center}
\end{figure}
It is quite well known that the PMD on the high
momentum side is affected by the energy conservation, the effect
of which on the PMD is expressed by the phase volume
$\bar{\rho}_{{\bm K}_{\mathrm{B}}^{\mathrm{A}}}$.
We use the PWIA, Eq.~(\ref{pmdpw}), to see the role of
$\bar{\rho}_{{\bm K}_{\mathrm{B}}^{\mathrm{A}}}$ clearly.
The solid line in Fig.~\ref{fig3} shows the result of the PWIA without
including the phase volume. More precisely, in this calculation
$\bar{\rho}_{{\bm K}_{\mathrm{B}}^{\mathrm{A}}}$ in Eq.~(\ref{pmdpw})
is replaced with
\beq
\bar{\rho}_0
\equiv
\dfrac
{
\displaystyle\int \bar{\rho}_{{\bm K}_{\mathrm{B}}^{\mathrm{A}}}
d {\bm K}_{\mathrm{B}}^{\mathrm{A}}
}
{
\displaystyle\int_{\bar{\rho}_{{\bm K}_{\mathrm{B}}^{\mathrm{A}}} \neq 0}
d {\bm K}_{\mathrm{B}}^{\mathrm{A}}.
}
\eeq
Then
$d\sigma^{\mathrm{PW}}/d\boldsymbol{K}_{\mathrm{B}}^{\mathrm{A}}$
is just proportional to the effective s.p. MD of $^{14}$O.
As a result, the solid line in
Fig.~\ref{fig3} has a purely symmetric shape with respect to
its centroid momentum located at ${P}_{\mathrm{B}z}^{\mathrm{A}}=0$.
If the phase volume
$\bar{\rho}_{{\bm K}_{\mathrm{B}}^{\mathrm{A}}}$
is taken into account as in Eq.~(\ref{pmdpw}),
the dashed line is obtained. One sees that
$\bar{\rho}_{{\bm K}_{\mathrm{B}}^{\mathrm{A}}}$
gives a quite sharp cut of the PMD on the high momentum side.
This results in the quite drastic shortening of the FWHM $\Gamma$,
which is used as a measure of $l$. The result shown in Fig.~\ref{fig3}
suggests that inclusion of
$\bar{\rho}_{{\bm K}_{\mathrm{B}}^{\mathrm{A}}}$, i.e., energy
and momentum conservation, is necessary to relate the PMD and $l$
properly. On the other hand, the peak height of the solid
line is almost the same as that of the dashed line.
The dotted line shows the result using
\beq
\bar{\rho}^{\rm av}
_{{K}_{\mathrm{B}z}^{\mathrm{A}}}
\equiv
\dfrac
{
\displaystyle\int
\bar{\rho}_{{\bm K}_{\mathrm{B}}^{\mathrm{A}}}
K_{\mathrm{B}b}^{\mathrm{A}}
d {K}_{\mathrm{B}b}^{\mathrm{A}}
}
{
\displaystyle\int_{\bar{\rho}_{{\bm K}_{\mathrm{B}}^{\mathrm{A}}} \neq 0}
K_{\mathrm{B}b}^{\mathrm{A}}
d {K}_{\mathrm{B}b}^{\mathrm{A}}
}
\eeq
instead of $\bar{\rho}_{{\bm K}_{\mathrm{B}}^{\mathrm{A}}}$ in
Eq.~(\ref{pmdpw}). Although the shape of the dotted line
is quite similar to that of the dashed line, the former
undershoots the latter at and below the centroid momentum
($\sim -80$~MeV/$c$). Thus, accurate treatment of
$\bar{\rho}_{{\bm K}_{\mathrm{B}}^{\mathrm{A}}}$ is found to be
important.

%%%%%%%%%%%%%%%%%%%%%%%
%%%  Figure 4
%%%%%%%%%%%%%%%%%%%%%%%
\begin{figure}[htbp]
\begin{center}
 \includegraphics[width=0.45\textwidth,clip]{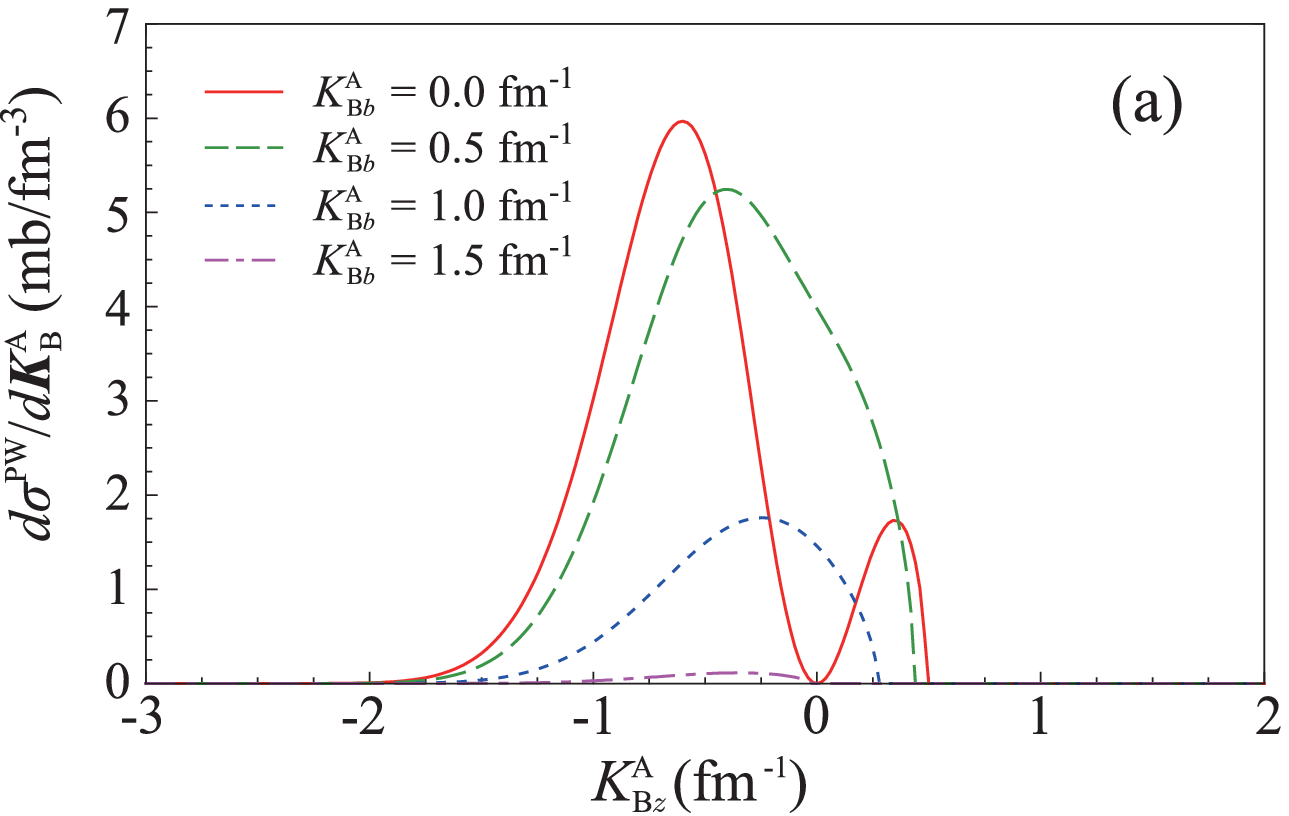}
 \includegraphics[width=0.45\textwidth,clip]{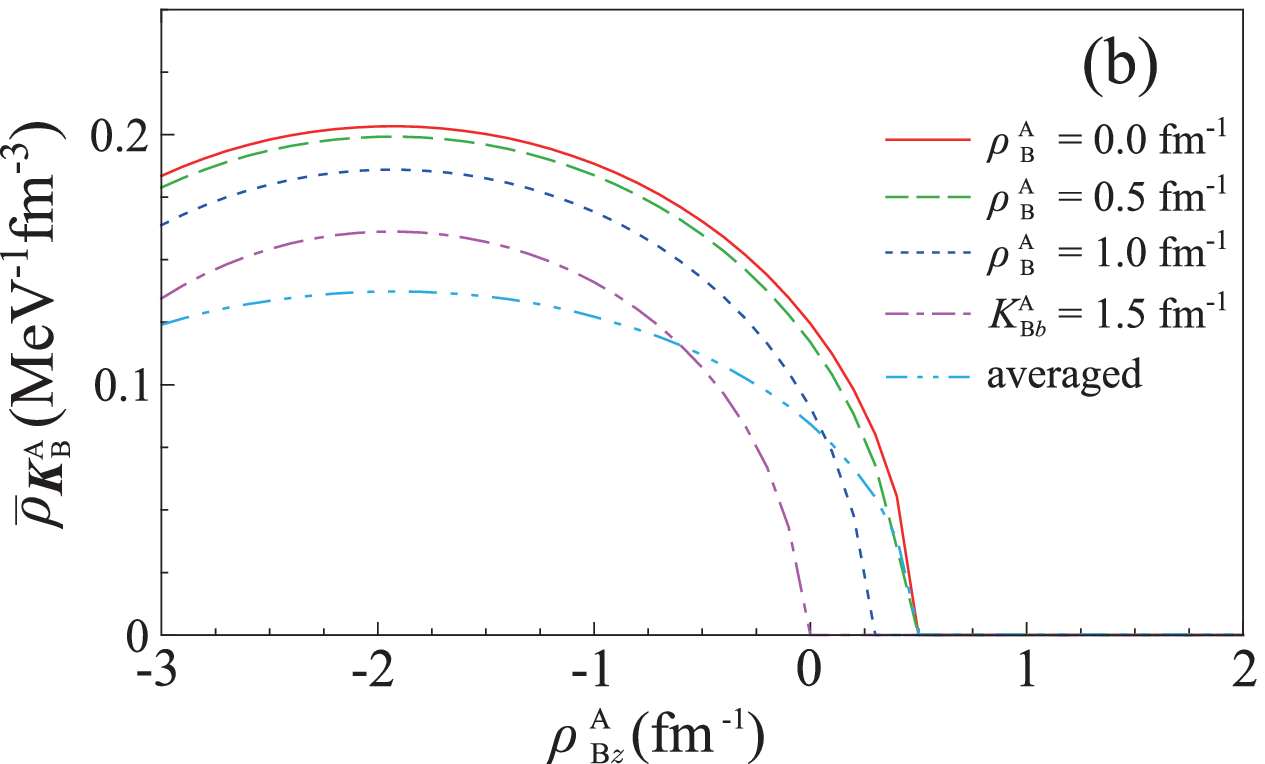}
 \includegraphics[width=0.45\textwidth,clip]{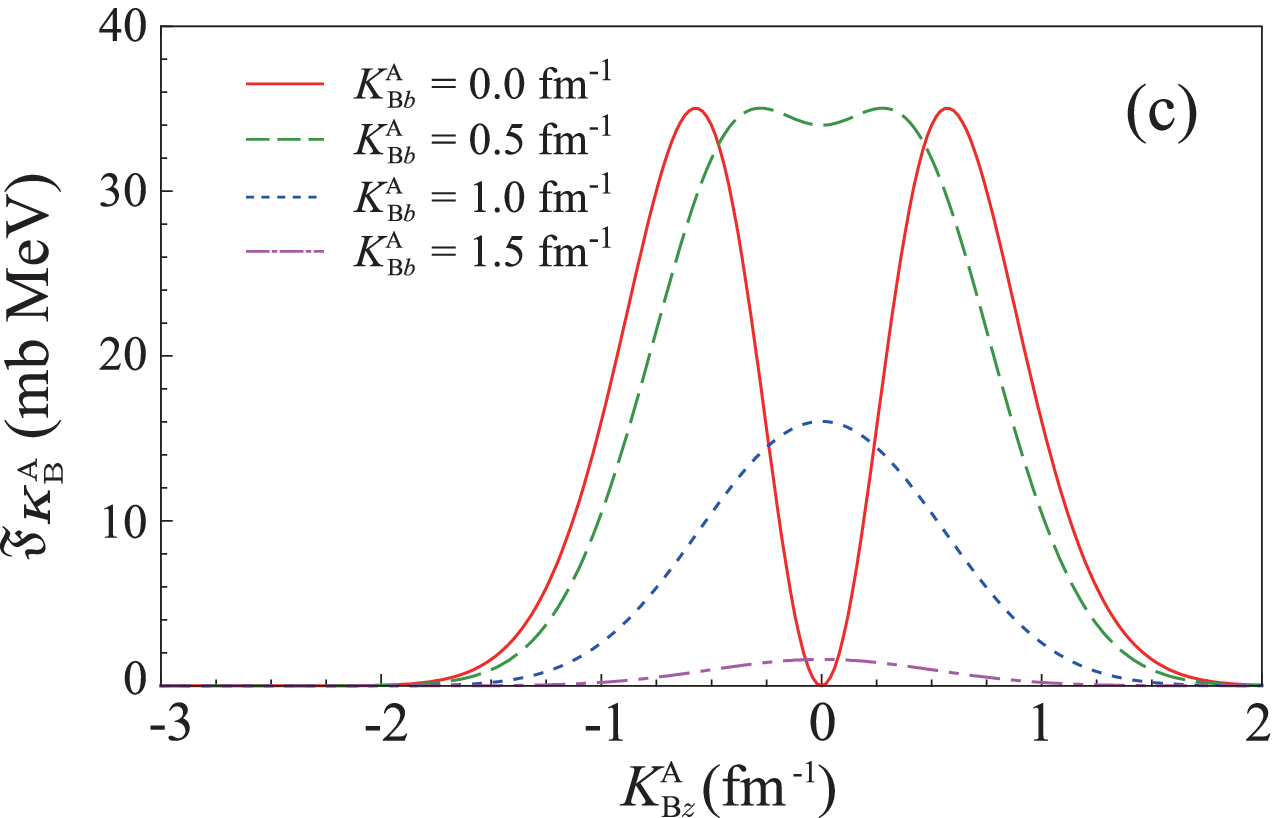}
 \caption{(Color online) (a) Momentum distribution
 calculated with the PWIA,
 $d\sigma^{\mathrm{PW}}/d\boldsymbol{K}_{\mathrm{B}}^{\mathrm{A}}$,
 (b) phase volume $\bar{\rho}_{{\bm K}_{\mathrm{B}}^{\mathrm{A}}}$,
 and (c) effective s.p. momentum distribution
 $\mathfrak{F}_{{\bm K}_{\mathrm{B}}^{\mathrm{A}}}$,
 for the $^{14}$O($p,pn$)$^{13}$O reaction residue at 100~MeV/nucleon
 in inverse kinematics.
 The solid, dashed, dotted, and dash-dotted lines in each panel
 correspond to $K_{\mathrm{B}b}^{\mathrm{A}}=0.0$, 0.5, 1.0, and
 1.5~fm$^{-1}$, respectively.
 The dash-dot-dotted line in (b) shows the averaged phase volume
 $\bar{\rho}^{\rm av}_{{K}_{\mathrm{B}z}^{\mathrm{A}}}$.
}
 \label{fig4}
\end{center}
\end{figure}
The role of $\bar{\rho}_{{\bm K}_{\mathrm{B}}^{\mathrm{A}}}$ is
seen more clearly in Fig.~\ref{fig4}.
The MD
$d\sigma^{\mathrm{PW}}/d\boldsymbol{K}_{\mathrm{B}}^{\mathrm{A}}$
is shown in Fig.~\ref{fig4}(a) and decomposed into
the phase volume
$\bar{\rho}_{{\bm K}_{\mathrm{B}}^{\mathrm{A}}}$ (Fig.~\ref{fig4}(b))
and the effective s.p. MD
$\mathfrak{F}_{{\bm K}_{\mathrm{B}}^{\mathrm{A}}}$
(Fig.~\ref{fig4}(c)). In each panel the solid,
dashed, dotted, and dash-dotted lines correspond to
$K_{\mathrm{B}b}^{\mathrm{A}}=0.0$, 0.5, 1.0, and 1.5~fm$^{-1}$,
respectively.
As shown in Fig.~\ref{fig4}(b),
$\bar{\rho}_{{\bm K}_{\mathrm{B}}^{\mathrm{A}}}$ shows clear
asymmetry with respect to
${K}_{\mathrm{B}z}^{\mathrm{A}}=0$. This behavior
can easily be understood by its nonrelativistic expression.
When $A\gg1$, one finds from Eq.~(\ref{pv}) that
i) $\bar{\rho}_{{\bm K}_{\mathrm{B}}^{\mathrm{A}}}^{\mathrm{NR}}$ has a maximum
at $K_{\mathrm{B}z}^{\mathrm{A}}=-K_{0}^{\mathrm{A}}$
and ii) $\bar{\rho}_{{\bm K}_{\mathrm{B}}^{\mathrm{A}}}^{\mathrm{NR}}=0$ at
$K_{\mathrm{B}z}^{\mathrm{A}}=-K_{0}^{\mathrm{A}}
\pm [2(K_{0}^{\mathrm{A}})^{2}
-(K_{\mathrm{B}b}^{\mathrm{A}})^{2}-\bar{S}_{N}]^{1/2}$.
The phase volume
$\bar{\rho}_{{\bm K}_{\mathrm{B}}^{\mathrm{A}}}$
plotted in Fig.~\ref{fig4}(b), i.e.,
calculated with the relativistic
kinematics, has the same features.

The effective s.p. MD at $K_{\mathrm{B}b}^{\mathrm{A}}=0$
shown by the solid line in Fig.~\ref{fig4}(c) has
a two peak structure and distributes up to
$|K_{\mathrm{B}z}^{\mathrm{A}}| \sim 2.0$~fm$^{-1}$, reflecting
the MD of the $0p3/2$ neutron inside $^{14}$O.
Since $\mathfrak{F}_{{\bm K}_{\mathrm{B}}^{\mathrm{A}}}$
is in fact independent of
the direction of ${\bm K}_{\mathrm{B}}^{\mathrm{A}}$,
the behavior of the other lines in Fig.~\ref{fig4}(c)
can be understood by that of the
solid line.
.
The MD of B is obtained by taking the product of
$\bar{\rho}_{{\bm K}_{\mathrm{B}}^{\mathrm{A}}}$ and
$\mathfrak{F}_{{\bm K}_{\mathrm{B}}^{\mathrm{A}}}$,
which strongly suppresses the high momentum side,
as shown in Fig.~\ref{fig4}(a).
One can find that the shape of the PMD (the dashed line in
Fig.~\ref{fig3}) is similar to the MD at
$K_{\mathrm{B}b}^{\mathrm{A}}=0.5$~fm$^{-1}$ (the dashed line
in Fig.~\ref{fig4}(a)).
This is because the MD at around this value of
$K_{\mathrm{B}b}^{\mathrm{A}}$ has the main contibution to the PMD.
It should be noted that the averaged phase volume
$\bar{\rho}^{\rm av}_{{K}_{\mathrm{B}z}^{\mathrm{A}}}$ shown by
the dash-dot-dotted line in Fig.~\ref{fig4}(b) quite well agrees
with the dashed line for ${K}_{\mathrm{B}z}^{\mathrm{A}} \ga 0$,
whereas the former undershoots the latter
for ${K}_{\mathrm{B}z}^{\mathrm{A}} \la 0$. This explains the reason for the
difference between the dashed and dotted lines in Fig.~\ref{fig3}.

%%Thus, the importance of the phase volume
%%$\bar{\rho}_{{\bm K}_{\mathrm{B}}^{\mathrm{A}}}$
%%in the description of the PMD is clarified.
More intuitive interpretation
of the role of $\bar{\rho}_{{\bm K}_{\mathrm{B}}^{\mathrm{A}}}$
can be given as follows.
Let us consider two cases: 1)
${\bm K}_{\mathrm{B}}^{\mathrm{A}}=K_{\mathrm{B}}^{\mathrm{A}}{\bm e}_z$
and 2)
${\bm K}_{\mathrm{B}}^{\mathrm{A}}=-K_{\mathrm{B}}^{\mathrm{A}}{\bm e}_z$.
Since B has the same energy in the two cases,
$E_1^{\mathrm{A}}+E_2^{\mathrm{A}}$ is fixed at a same value because
of the energy conservation. In a more rough estimation,
the kinetic energy of B, $T_{\rm B}$, is considered to be
negligibly small. Then one finds
\beq
T_1^{\mathrm{A}}+T_2^{\mathrm{A}}\approx T_0^{\mathrm{A}}-S_N.
\label{t12}
\eeq
On the other hand, the momentum conservation restricts the sum of
$K_{1z}^{\mathrm{A}}$ and $K_{2z}^{\mathrm{A}}$
to be
$-K_{0}^{\mathrm{A}}-K_{\mathrm{B}}^{\mathrm{A}}$ in case 1)
and
$-K_{0}^{\mathrm{A}}+K_{\mathrm{B}}^{\mathrm{A}}$ in case 2);
note that
${\bm K}_{0}^{\mathrm{A}}=-K_{0}^{\mathrm{A}}{\bm e}_z$
in the A-rest frame.
$|K_{1z}^{\mathrm{A}} + K_{2z}^{\mathrm{A}}|$ in the former is much
larger than in the latter. Under the condition of Eq.~(\ref{t12}),
it is difficult to satisfy the momentum conservation in case 1).

When $S_N$ is small, the cutoff momentum for
$\bar{\rho}_{{\bm K}_{\mathrm{B}}^{\mathrm{A}}}$
becomes larger. In addition to that, the effective s.p. MD
becomes narrower. These two make the effect of
$\bar{\rho}_{{\bm K}_{\mathrm{B}}^{\mathrm{A}}}$
on the MD smaller. This is the reason for the difference in
the PMD of $^{13}$O (solid line) and $^{13}$N (dashed line)
on the high momentum side shown in Fig.~\ref{fig2}.
One can also expect a smaller effect of
$\bar{\rho}_{{\bm K}_{\mathrm{B}}^{\mathrm{A}}}$
at higher incident energy. We will return to this point in
Sec.~\ref{sec35}.

\subsection{Distortion effect on the low momentum side}
\label{sec34}

One sees by comparing the solid line in Fig.~\ref{fig2} and
the dashed line in Fig.~\ref{fig3} that the distortion
generates a well-developed low momentum tail in the PMD.
In the present eikonal DWIA model, as mentioned, the effect
of distortion is  aggregated into the distorted-wave factor
$F_{K_{0}K_{1}K_{2}}$; its form in inverse kinematics is
given in Eq.~(\ref{finv}).
One sees from Eq.~(\ref{finv}) that
the imaginary part $W_i$ of each optical potential gives
a reduction of the amplitude of $F_{K_{0}K_{1}K_{2}}$, i.e., absorption.
On the other hand, the real part $V_i$ generates an effective
momentum $\Delta{\bm K}_{i}$ for each particle;
$\Delta{\bm K}_{0}$ is anti-parallel with the $z$-axis
and $\Delta{\bm K}_{1}$ and $\Delta{\bm K}_{2}$
are parallel with the $z$-axis. One may interpret
$\Delta{\bm K}_{i}$ as the momentum shift of a nucleon inside
an attractive potential due to the local energy conservation.

Thus, the exponent $-i K_{Nz}z$ in Eq.~(\ref{tmat})
is effectively changed as
\bea
-i K_{Nz}z
&\to&
-i K_{Nz}z
+ i(-\Delta K_{0} + \Delta K_{1} + \Delta K_{2})z
\non \\
&\approx&
i (K_{\mathrm{B}z}^{\mathrm{A}} - \Delta K_{0} + \Delta K_{1} + \Delta K_{2})z,
\eea
where use has been made of
$\boldsymbol{K}_{N}\approx - \boldsymbol{K}_{\mathrm{B}}^{\mathrm{A}}$.
It is found that $\Delta K_{0}$ is quite small
because the kinetic energy of particle 0 is fixed at 100~MeV.
On the other hand, particles 1 and 2 are allowed to have quite lower
energies, for which the distortion effect is larger.
Then $K_{\mathrm{B}z}^{\mathrm{A}}$ is effectively shifted toward
the $+z$ direction by $V_1$ and $V_2$.
In other words, the PMD
$d\sigma/d {K}_{\mathrm{B}z}^{\mathrm{A}}$ probes the nucleon inside
the nucleus A that has the longitudinal momentum around
$|K_{\mathrm{B}z}^{\mathrm{A}} + \Delta K_{1} + \Delta K_{2}|$.

%%%%%%%%%%%%%%%%%%%%%%%
%%%  Figure 5
%%%%%%%%%%%%%%%%%%%%%%%
\begin{figure}[htbp]
\begin{center}
 \includegraphics[width=0.45\textwidth,clip]{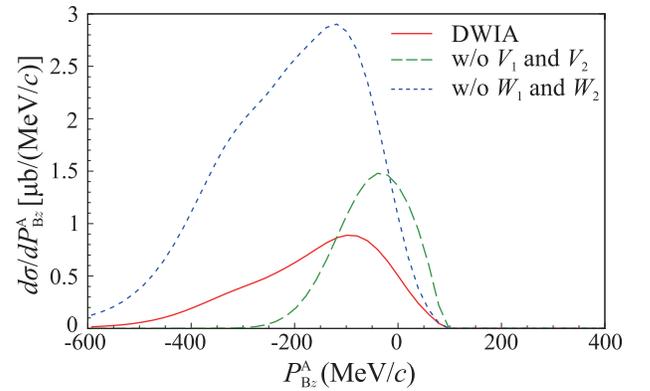}
 \caption{(Color online) Parallel momentum distribution of
 $^{13}$O for $^{14}$O($p,pn$)$^{13}$O at 100~MeV/nucleon
 in inverse kinematics. The solid line is the DWIA result,
 whereas the dashed (dotted) line represents
 the result calculated with $V_i=0$ ($W_i=0$) for particles 1 and 2.
}
 \label{fig5}
\end{center}
\end{figure}
To see this, we show in Fig.~\ref{fig5} the PMD of $^{13}$O for
$^{14}$O($p,pn$)$^{13}$O at 100~MeV/nucleon in inverse kinematics,
calculated with the DWIA putting $V_i=0$ (dashed line) and $W_i=0$ (dotted line)
for particles 1 and 2. For comparison we show by the solid line the
DWIA result, which is the same as in Fig.~\ref{fig2}.
As clearly shown, $V_1$ and $V_2$ generate the low momentum
tail. It should be remarked that the shift of the PMD
towards the low momentum side by $V_1$ and $V_2$ affects also the
height of the peak of the PMD. On the other hand, $V_1$ and $V_2$
change the integrated cross section by only about 5\%.

In the actual calculation, as mentioned, the energy dependence
of $U_1$ and $U_2$ is explicitly taken into account. It is found
that the qualitative feature of the result shown in Fig.~\ref{fig5}
does not change even if $U_1$ and $U_2$ are evaluated at a
fixed energy, $E_{\rm fix}$.
Quantitatively, however, this treatment affects the result. The height
of the peak changes by about 60\%, depending on the value of $E_{\rm fix}$.

It is found that the momentum shift appears also in
$^{14}$O($p,2p$)$^{13}$N but the effect is quite small. This
can be understood by the difference in $S_N$. Equation~(\ref{t12})
shows that the kinetic energies of particles 1 and 2 are more
severely restricted when $S_N$ is large. Then the distortion effects
due to $V_1$ and $V_2$ become more important.

\subsection{Results at 200~MeV/nucleon}
\label{sec35}

%%%%%%%%%%%%%%%%%%%%%%%
%%%  Figure 6
%%%%%%%%%%%%%%%%%%%%%%%
\begin{figure}[htbp]
\begin{center}
 \includegraphics[width=0.45\textwidth,clip]{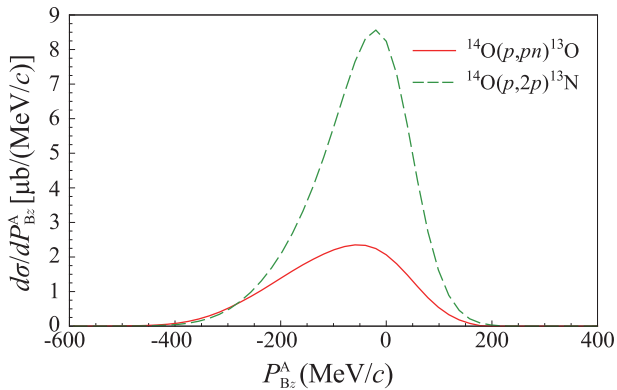}
 \caption{(Color online) Same as Fig.~\ref{fig2} but at
 200~MeV/nucleon.
}
 \label{fig6}
\end{center}
\end{figure}
Both the phase volume effect and the distortion effect discussed
above will become less important as the incident energy increases.
We show in Fig.~\ref{fig6} the PMD of the
$^{14}$O($p,pN$)$^{13}$O at 200~MeV/nucleon. The meaning of the
lines is the same as in Fig.~\ref{fig2}. The two lines agree with
each other around the tail on the high momentum side. This
suggests that the phase volume effect becomes small, though not
negligible.
On the other hand, both results still show somewhat large
asymmetry, i.e., $A_\Gamma$, indicating the importance of the
distortion for particles 1 and 2 at this incident energy.
The features of the results are summarized in Table~\ref{tab2}.
%
%%%%%%%%%%%%%%%%%%%%%%%
%%%  Table 2
%%%%%%%%%%%%%%%%%%%%%%%
\begin{table}[hptb]
\caption{Same as Table~\ref{tab1} but at 200~MeV/nucleon; the result
for $^{31}$Ne($p,pn$)$^{30}$Ne is also shown.}
\label{tab2}
\begin{tabular}{cccccc}
\hline
\hline
nucleus  & $P_{\mathrm{cen}}$~(MeV/$c$) & $\Gamma$~(MeV/$c$)
         & $\Gamma_{\mathrm{L}}$~(MeV/$c$)
         & $\Gamma_{\mathrm{H}}$~(MeV/$c$) & $A_{\Gamma}$ \\
\hline
$^{13}$O  & $-53$  & 278 & 168 & 110 & 1.53 \\
$^{13}$N  & $-20$  & 191 & 113 &  78 & 1.45 \\
$^{30}$Ne & $-0.7$ &  52 &  32 &  30 & 1.08 \\
\hline
\hline
\end{tabular}
\end{table}
%

%%%%%%%%%%%%%%%%%%%%%%%
%%%  Figure 7
%%%%%%%%%%%%%%%%%%%%%%%
\begin{figure}[htbp]
\begin{center}
 \includegraphics[width=0.45\textwidth,clip]{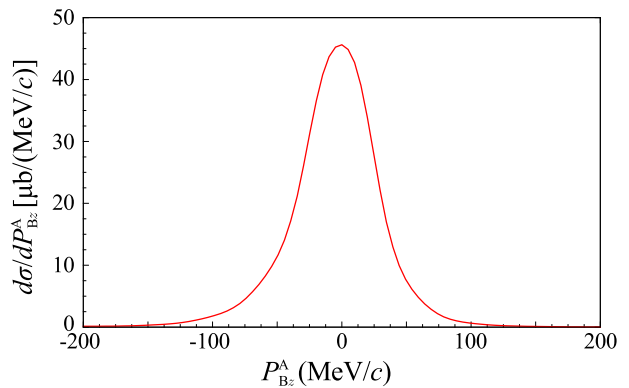}
 \caption{(Color online) Same as Fig.~\ref{fig6} but for
 $^{31}$Ne($p,pn$)$^{30}$Ne.
}
 \label{fig7}
\end{center}
\end{figure}
When $S_N$ is even smaller, the PMD becomes almost symmetric,
as shown in Figure~\ref{fig7}, in which the PMD of $^{30}$Ne
for $^{31}$Ne($p,pn$)$^{30}$Ne at 200~MeV/nucleon in inverse
kinematics is plotted. We assume that the $1p3/2$ neutron with
$S_N=0.15$~MeV is knocked out. The most important feature of
this reaction is that the $1p3/2$ neutron has a very narrow
s.p. MD because of its small $S_N$.
Within the range corresponding to the s.p. MD,
$K_{\mathrm{B}z}^{\mathrm{A}}$ essentially gives no effect
on the kinematics of the three-body system.
In addition to that, since the $1p3/2$ neutron forms a halo,
the main contribution to the reduced transition amplitude Eq.~(\ref{tmat})
comes from the surface region of the nucleus, which significantly
suppresses the distortion effect.
It should be noted that this is the case with only a
very weakly-bound nucleus;
for the $^{14}$O($p,2p$)$^{13}$O process corresponding to
a quite small value of $S_N$ (4.63~MeV), the distortion effect
still exists as shown in Fig.~\ref{fig6}.
The PMD for a knockout process of
a very weakly-bound nucleon at around 200~MeV/nucleon
shows, therefore, a symmetric shape rather exceptionally.

\section{Summary}
\label{sec4}

We have investigated the PMD of the residual nuclei of the
$^{14}$O($p,pn$)$^{13}$O and $^{14}$O($p,2p$)$^{13}$N reactions
at 100 and 200~MeV/nucleon in inverse kinematics.
An eikonal DWIA model was adopted, which was shown to reproduce
the TDX data of the $^{12}$C($p,2p$)$^{11}$B at 392~MeV very well.
The PMD of both $^{13}$O and $^{13}$N have an asymmetric shape
at 100~MeV/nucleon. The high momentum side steeply falls, whereas
a well-developed tail exists on the low momentum side.

The former is found to be due to the phase volume effect reflecting
the energy and momentum conservation. We have clarified
how the phase volume affects the s.p. MD of the nucleon inside
$^{14}$O in detail by using PWIA.
The width $\Gamma$ of the PMD is much smaller than that of the s.p. MD
by the phase volume effect. This should be remarked because $\Gamma$
is used as a measure of the s.p. orbital angular momentum $l$.
On the other hand, the phase volume does not change the peak height
of the PMD.
The phase volume effect becomes less important when $S_N$ is small
because 1) the cutoff momentum of the phase volume on the high momentum
side is large and 2) the width of the s.p. MD is small.

The latter, the tail of the PMD on the low momentum side, is found
to be due to the momentum shift of the outgoing two nucleons
inside an attractive potential caused by the residual nucleus.
Consequently, the PMD $d\sigma/d {K}_{\mathrm{B}z}^{\mathrm{A}}$
probes the nucleon inside the nucleus A
having the longitudinal momentum
$|K_{\mathrm{B}z}^{\mathrm{A}} + \Delta K_{\mathrm{eff}}|$,
where $\Delta K_{\mathrm{eff}}$ ($>0$) is the effective momentum
due to the distortion effect. It should be noted that
$\Delta K_{\mathrm{eff}}$ gives a somewhat large reduction of
the peak height of the PMD, which is a key quantity to determine the
spectroscopic factor ${\cal S}$. The momentum shift has a quite
small effect ($\sim 5$\%) on the integrated cross section.

We found that at 200~MeV/nucleon the phase volume effect becomes less
important, whereas the distortion effect still exists. For
the $^{31}$Ne($p,pn$)$^{30}$Ne reaction at 200~MeV/nucleon,
exceptionally, the PMD has an almost symmetric shape. This is
because of the very small value (0.15~MeV) of $S_N$ in this case.
It should be remarked that the small distortion effect is
due to the halo structure of $^{31}$Ne; the contribution of the
nuclear interior region, where distorting potentials are large,
to the ($p,pN$) transition amplitude is almost negligible.

For a quantitative comparison with experimental data of a PMD,
on the low momentum side in particular, use of non-eikonal
scattering wave functions will be necessary. Extension of
the present DWIA framework to knockout reactions by a nucleus
will also be important for discussing various experimental data
of nucleon removal processes measured so far.

\section*{Acknowledgments}

The authors thank T.~Noro for providing them with updated
experimental data before publication.
One of the authors (K.O.) thanks C.~A.~Bertulani and P.~Capel
for helpful discussions. This work was supported in part
by Grant-in-Aid of the Japan Society for the
Promotion of Science (Grant No. 25400255 and No. 15J01392)
and by the ImPACT Program of the Council for Science, Technology and
Innovation (Cabinet Office, Government of Japan).

%%--------------------------------------------------------------------%%
%%                           References                               %%
%%--------------------------------------------------------------------%%

\end{document}